# Transcranial photoacoustic computed tomography of human brain function

Yang Zhang[1†], Shuai Na[1†], Karteekeya Sastry[6†], Jonathan J. Russin[2,3,4†], Peng Hu[1], Li Lin[1], Xin Tong[1], Kay B. Jann[5], Danny J. Wang[5], Charles Y. Liu[2,3,4*], Lihong V. Wang[1,6,*]

[1] Caltech Optical Imaging Laboratory, Andrew and Peggy Cherng Department of Medical Engineering, California Institute of Technology, 1200 East California Boulevard, Pasadena, CA 91125, USA.

[2] Department of Neurological Surgery, Keck School of Medicine, University of Southern California, Los Angeles, CA 90033, USA.

[3] Neurorestoration Center, Keck School of Medicine, University of Southern California, Los Angeles, CA 90033, USA.

[4] Rancho Los Amigos National Rehabilitation Center, Downey, CA 90242, USA.

[5] Laboratory of Functional MRI Technology, Stevens Neuroimaging and Informatics Institute, Keck School of Medicine, University of Southern California, Los Angeles, CA 90033, USA.

[6] Caltech Optical Imaging Laboratory, Department of Electrical Engineering, California Institute of Technology, 1200 East California Boulevard, Pasadena, CA 91125, USA.

[†]These authors contributed equally to this work: Yang Zhang, Shuai Na, Karteekeya Sastry, Jonathan J. Russin.

*Corresponding author. C. Y. Liu (cliu@usc.edu), L. V. Wang (LVW@caltech.edu)

## Abstract

Herein we report the first in-human transcranial imaging of brain function using photoacoustic computed tomography. Functional responses to benchmark motor tasks were imaged on both the skull-less and the skull-intact hemispheres of a hemicraniectomy patient. The observed brain responses in these preliminary results demonstrate the potential of photoacoustic computed tomography for achieving transcranial functional imaging.

## Introduction

Although several imaging modalities exist for functional brain imaging — blood oxygen level-dependent (BOLD) functional magnetic resonance imaging (fMRI)[1], electroencephalography (EEG)[2], functional near-infrared spectroscopy (fNIRS)[3], and functional ultrasound imaging (fUS)[4,5] — each of these methods faces certain limitations. BOLD fMRI is the gold standard for functional brain imaging[1]; however, it exhibits a nonlinear relationship with the hemoglobin concentration. EEG measures the electrophysiological activities at a high temporal resolution, but it suffers from poor spatial resolution[2]. fNIRS has higher temporal resolution than fMRI, but poorer spatial resolution[3]. fUS has been reported for intraoperative cortical functional mapping[4], whereas transcranial imaging of human cerebral vasculature at microscopic resolution requires the use of contrast agents[6].

Photoacoustic computed tomography (PACT) is a complementary technique for functional brain imaging[7,8]. Compared with fMRI, it has the advantage of linear detection of oxyhemoglobin (HbO$_2$) and deoxyhemoglobin (HbR) concentrations with minimal extravascular interference. High spatiotemporal resolution can be achieved using a large number of ultrasonic detection channels with a sufficient detection view. Recently, the potential of PACT to image human brain function has been demonstrated[9]. This prior report leveraged a unique clinical population of hemicraniectomy patients, and it demonstrated the potential of functional PACT (fPACT) in the absence of the skull. The human skull presents a major obstacle for transcranial fPACT[10,11]. The clinical relevance of fPACT for human brain imaging is critically linked to the ability to overcome this obstacle.

Herein, we describe the first in-human transcranial imaging of brain function using fPACT. Adequate depth of penetration was achieved using a 1064 nm laser, and the photoacoustic signals were detected using a synthetic hemispherical array. The system was operated in functional mode to form a volumetric image with a scan time of 2 s and in structural mode to form a volumetric angiography image in 10 s. We imaged both the skull-less and skull-intact hemispheres of a hemicraniectomy patient performing finger tapping (FT) and tongue tapping (TT) tasks. The results showed functional responses on the skull-intact hemisphere but appeared to be scattered, presumably due to skull aberration. This preliminary detection of functional responses through the skull demonstrates the potential of fPACT for transcranial functional imaging.

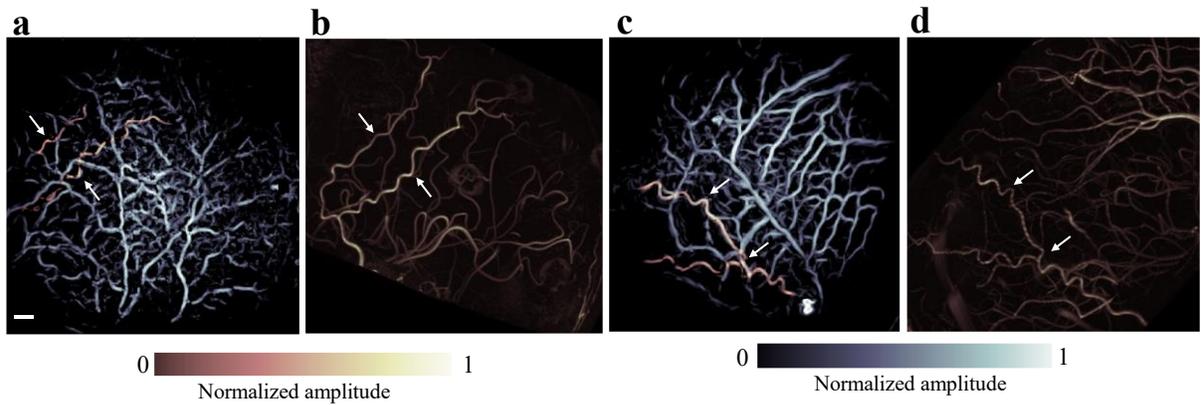

**Fig. 1 | PACT angiography and MRA of the same brain. a-b**, PACT angiography image and MRA image of the left skull-less side of the head, respectively. **c-d**, PACT angiography image and MRA image of the right skull-intact side of the head, respectively. White arrows indicate the locations of superficial scalp arteries (pink color) used for co-registration between the PACT and MRA images. Scale bar: 1 cm.

## Results

**Brain Angiography.** The PACT system has been described previously[9]. It was first operated in structural mode to acquire angiographic images of both the left skull-less side and the right skull-intact side of a hemicraniectomy patient. The corresponding structural angiography images are shown in Figs. 1a and 1c, respectively. Images were also acquired using 7 T time-of-flight magnetic resonance angiography (MRA) as the gold standard (Figs. 1b and 1d). The images from the two modalities were co-registered using the superficial arteries indicated by pink color and white arrows in Figs. 1a-d. Note that PACT shows both arteries and veins, whereas MRA shows the arteries only.

**Mapping motor functions through the skull.** We imaged motor tasks in the left hemicraniectomy patient on both sides of the head using fPACT and 7 T fMRI. We employed two motor tasks — FT and TT — for the functional study. The stimulation paradigms and protocols are described in Methods. Functional responses to these tasks were extracted using the general linear model (GLM)[12]. The extracted functional responses were overlaid on the T1-weighted and MRA images after co-registering the MRA images with the PACT images. The functional responses imaged by fMRI and fPACT to the two motor tasks are shown in Fig. 2. For the right-hand FT task, we compare the functional activation detected on the left skull-less side by fPACT (Fig. 2b) with that detected by fMRI (Fig. 2a), and we find an acceptable agreement. Similarly, for the left-hand FT task, we compare the functional activations detected by fMRI (Fig. 2c) and fPACT (Fig. 2d).

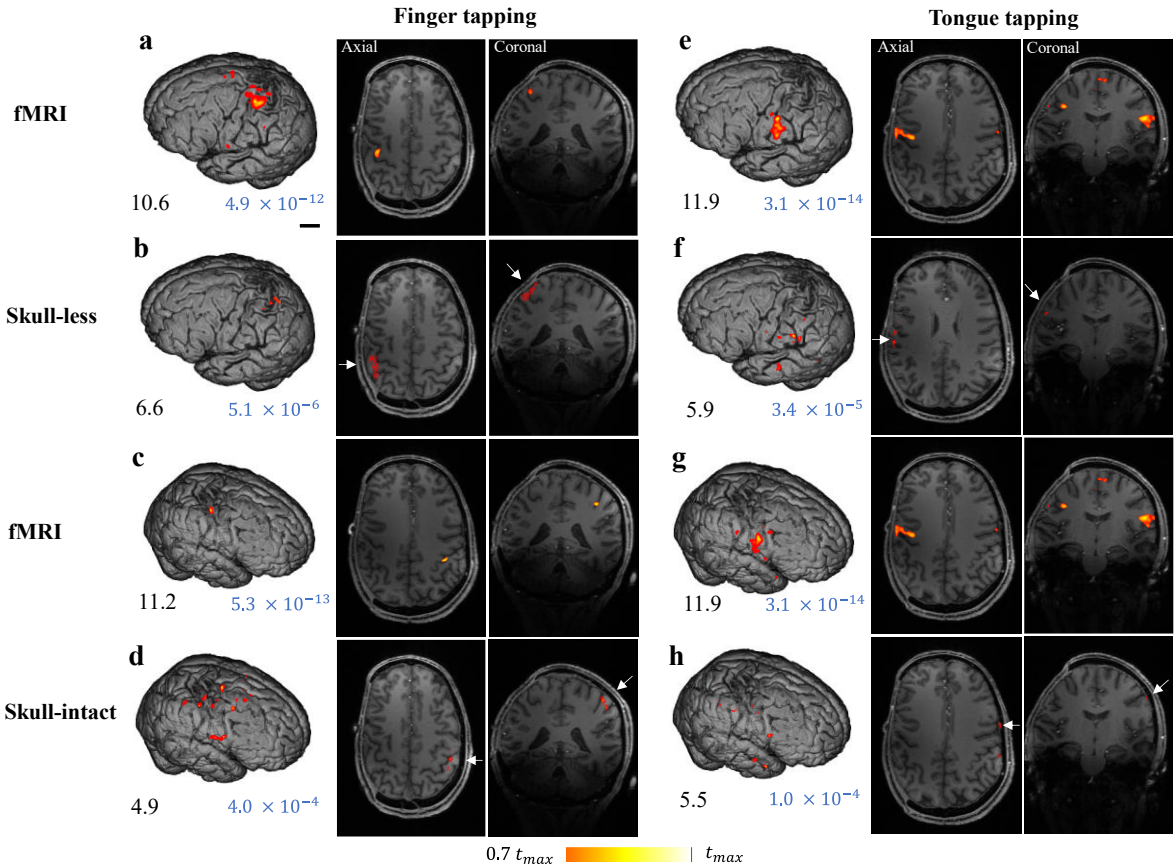

**Fig. 2 | fPACT and 7 T fMRI results for motor tasks.** Functional responses to right-hand FT (**a**: fMRI, **b**: left-hemisphere skull-less fPACT), left-hand FT (**c**: fMRI, **d**: right-hemisphere skull-intact fPACT), and TT (**e**: fMRI — left panel shows the left side of the brain, **f**: left-hemisphere skull-less fPACT, **g**: fMRI — left panel shows the right side of the brain, **h**: right-hemisphere skull-intact fPACT) were imaged. Functional responses displayed on the cortex (left column) represent the maximum amplitude projection of the responses. Functional responses are also displayed on axial (middle column) and coronal (right column) slices passing through the activations. For FT (**a-d**), we choose the same axial and coronal slices for display in all four images. For TT on the left skull-less side (**e,f**), we choose slices that are within 5 mm of each other. For TT on the right skull-intact side (**g,h**), we choose the same axial and coronal slices. However, these activations do not overlap spatially. In each functional map, we show the regions thresholded at 70% of the maximum t-value ($t_{max}$), listed as the first value below each cortical map. The p-value corresponding to 70% of the maximum t-value is shown below the cortical map (one-sided Student's t-test). White arrows indicate the activated regions in fPACT. Scale bar: 2 cm.

Although the functional activation detected by fPACT is highly scattered, it does overlap with that detected by fMRI. The percentage changes in the fPACT signals during FT on the left skull-less and right skull-intact sides are shown to be comparable (Supplementary Fig. 2). For the TT task, the fMRI and fPACT activations on the skull-less side (Figs. 2e and 2f respectively) are within a distance of 5 mm of each other. Accordingly, we choose axial (middle column) and coronal slices (right column) that are within 5 mm of each other in Figs. 2e and 2f. Finally, we compare the fMRI and fPACT activations on the skull-intact side from the TT task (Figs. 2g and 2h respectively). Although functional activation is detected by fPACT (Fig. 2h), it is highly scattered and does not overlap spatially with the fMRI activation (Fig. 2g), again presumably due to skull aberration.

## Discussion

This study reports preliminary results for the first in-human transcranial functional brain imaging using the fPACT technique, which was presented on Jan. 22, 2022 in an overview talk at Photonics West[13]. Prior work has demonstrated that human brain function imaged by fPACT on the skull-less side of the brain has a strong spatial correspondence with that imaged by fMRI[9]. Here, we imaged functional brain responses to two motor tasks on both the skull-intact side and the skull-less side of a hemicraniectomy patient. On the skull-less side, we observed good spatial agreement between fPACT and fMRI activations, as expected. In contrast, on the skull-intact side, the fPACT activations are highly scattered and are in only partial agreement with fMRI activations. The discrepancies are attributable to the acoustic effect of the skull.

One of the significant hurdles in achieving transcranial fPACT is the presence of the skull. The skull induces optical and acoustic attenuation as well as acoustic aberration, which degrade the PACT image quality. Prior studies reported that an adult human skull transmits ~50% light intensity at 1064 nm and ~20% acoustic pressure[14]. The effect of the skull is reflected in the calculated p-values in Fig. 2, which are higher for the skull-intact side than for the skull-less side. The functional activations on the skull-intact side are more scattered than those on the skull-less side. We investigated the effects of the acoustic aberration caused by the skull on cortical vessels through phantom experiments (see Supplementary Fig. 3). In the absence of the skull, we could observe a clear boundary of the phantom target. In contrast, in the presence the skull, we could only observe a blurred branch-like target. The contrast-to-noise ratios (CNRs) of the target are 25.2 and 229.3 with and without the skull, respectively. Despite the blurring and the drop in CNR, the image with the skull still bears a distinct signature of the target being imaged. This experiment explains why we can detect functional changes through the skull, and also helps explain the skull-induced positional differences in the activations imaged by fPACT. We are investigating deaberration methods to improve the spatial resolution.

Although the preliminary results presented in this study require further validation with a larger subject population, the detection of functional responses through the skull to benchmark motor tasks observed in this study demonstrates the potential of PACT for transcranial functional brain imaging.

**Methods**

**Data acquisition.** We use a 1064 nm laser (Litron, pulse repetition frequency (PRF): 20 Hz, maximum pulse energy: ~2.5 J, pulse width: 4-7 ns) and an engineered diffuser (EDC-80, RPC Photonics Inc), mounted on the bottom of the bowl-shaped array housing, to deliver the diffuse beam to the target. We detect the photoacoustic signals using a bowl-shaped ultrasonic transducer array (Imasonic Inc.). To synchronize the laser with the data acquisition card (DAQ), we use the laser to send a trigger signal to the control and synchronization module which then triggers the DAQs. Finally, we stream the data to the computer via USB 3.0. The schematic of the system can be found in Supplementary Fig. 1.

We performed the fPACT scan of the subject in a supine position with a plastic film to separate the human head from the deuterium oxide ($D_2O$, Isowater Corp.) inside the array bowl. The gap between the head and plastic film was filled with a small amount of water to maintain optimal acoustic coupling for fPACT scanning. The system was operated in structural mode to obtain a volumetric angiography image in 10 seconds, and was operated in functional mode to obtain a volumetric image in 2 seconds, thus achieving a functional volumetric imaging rate of 0.5 Hz. The subject performed finger-tapping (unilateral thumb-pinky) to stimulate one side of the brain and tongue-tapping (tapping the roots of the upper incisors with the tongue in a closed mouth at a frequency of about 1 Hz) to stimulate both sides of the brain. We first performed the fPACT scans on the left skull-less side for right-hand FT and TT tasks. Then, we performed fPACT on the right skull-intact side for left-hand FT and TT tasks. The stimulation paradigm consisted of a 30-s control state at the beginning for baseline sampling followed by seven cycles of 60-s period (30-s active state and 30-s control state). During each stimulation phase, the subject repeated the motor task at a constant pace during the active state and relaxed during the control state. MRI data were acquired using a 7 T Siemens Terra scanner (Siemens Medical Solutions) and a Nova Medical 32-channel head coil (Siemens Medical Solutions) with the same stimulation paradigm.

**Image processing.** The image reconstruction of PACT was based on a dual-speed-of-sound universal back-projection (UBP) algorithm implemented in MATLAB, C++, and Python to account for acoustic inhomogeneities between the head and the coupling fluid ($D_2O$)[15]. The reconstructed volumetric images were motion-corrected and applied to a GLM to extract the activated regions of the functional map.

**Imaging protocols**

All human imaging experiments were performed with the relevant guidelines and regulations approved by the Institutional Review Board of the California Institute of Technology (Caltech). The human brain experiments were performed in a dedicated imaging room. Written informed consents were obtained from all the participants according to the study protocols.

**Data availability**

The data that support the findings of this study are provided within the paper and its supplementary material.

**Code availability**

The reconstruction algorithm and data processing methods can be found in Methods. The reconstruction code is not publicly available because it is proprietary and is used in licensed technologies.

**Acknowledgments**

We thank former lab members Konstantin Maslov, Xiaoyun Yuan, and Junhui Shi for their contribution towards building the system. This work was sponsored by the United States National Institutes of Health (NIH) grants U01 NS099717 (BRAIN Initiative), S10-OD025312, R01 NS102213, and R35 CA220436.

**Contributions**

L.V.W., C.Y.L., Y.Z., S.N. and J.J.R. designed the study. L.L., S.N., Y.Z., and X.T. built and modified the system hardware. S.N. and Y.Z. developed and modified the control program. P.H., and S.N. developed the structural and functional reconstruction algorithms. Y.Z., K.S., S.N., and J.J.R. performed the PACT experiments. K.B.J. performed the MRI experiments. K.S., Y.Z., S.N., C.Y.L., D.J.W. and L.V.W. analyzed and interpreted the data. J.J.R. and C.Y.L. recruited the participants. Y.Z., S.N., K.S., and J.J.R. wrote the manuscript with input from all authors. L.V.W. revised the manuscript. L.V.W., C.Y.L. and D.J.W. supervised the study.

**Competing interests**

L.V.W. has a financial interest in Microphotoacoustics Inc., CalPACT LLC, and Union Photoacoustic Technologies Ltd., which, however, did not support this work.

**Supplementary figures**

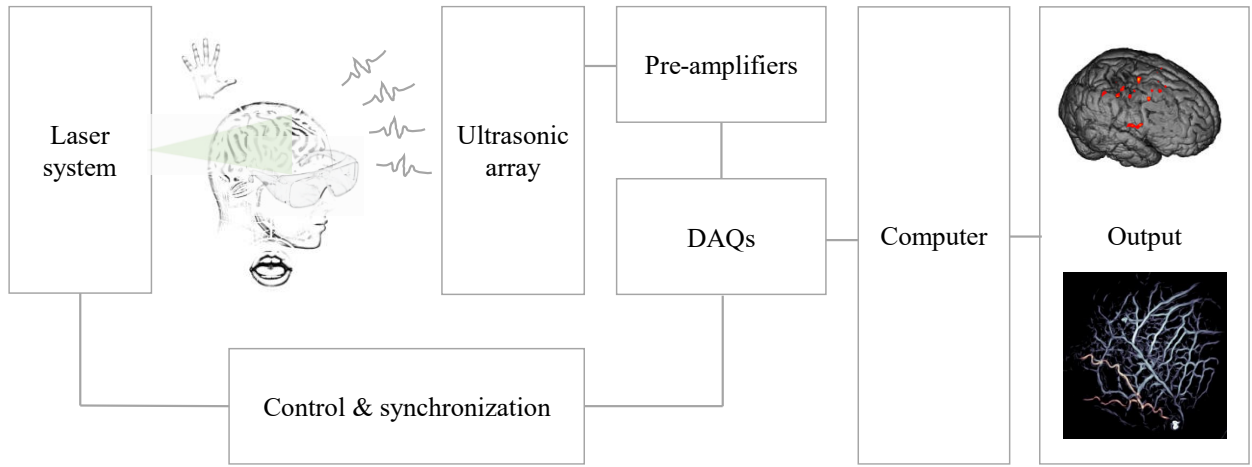

**Supplementary Fig. 1 | System schematic.** A 1064 nm laser delivers light through an engineered diffuser to the brain. A control and synchronization module synchronizes the light delivery and photoacoustic signal detection. An ultrasound array detects photoacoustic signals, which are amplified and digitized by the pre-amplifiers and the data acquisition cards (DAQs), respectively. The data are then streamed to a computer to form volumetric images. Finally, functional activation maps for the finger tapping and tongue tapping tasks are generated and overlaid onto the MRI images.

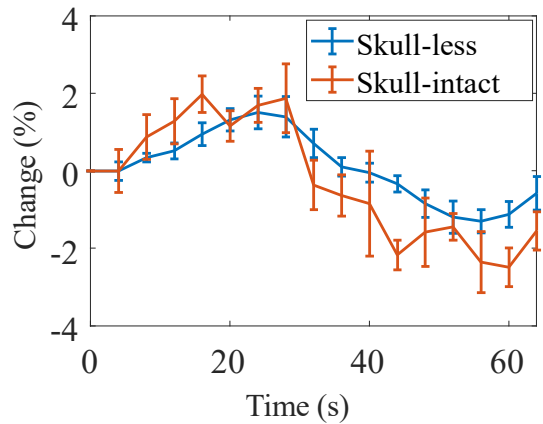

**Supplementary Fig. 2 | Fractional signal changes observed by fPACT.** The fractional signal changes on both sides show an increase in the active period (0-30 s) and a decrease in the control period (30-60 s). The changes on the skull-intact side of the brain have higher standard errors than those on the skull-less side.

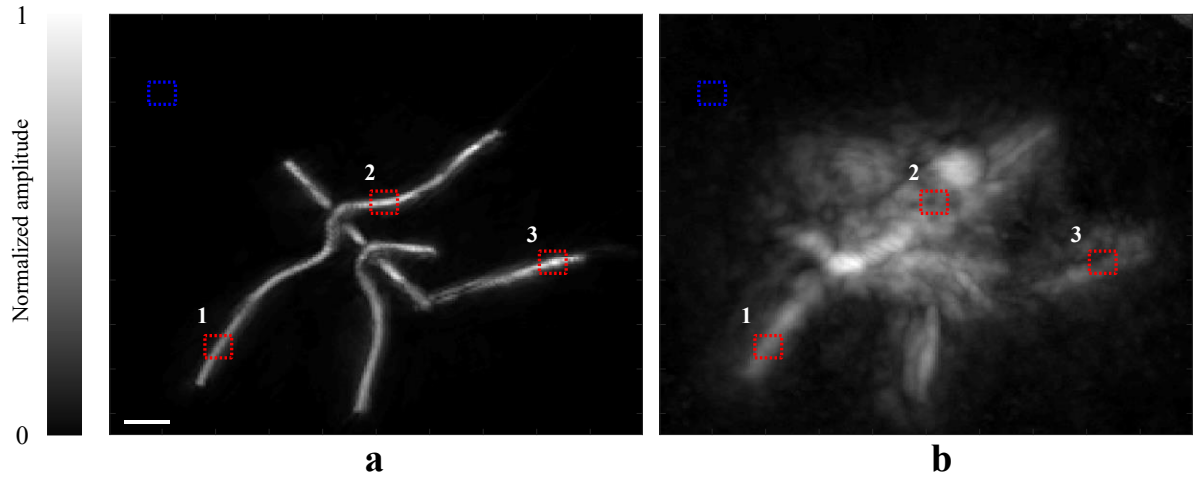

**Supplementary Fig. 3 | Phantom experimental images.** A light absorbing phantom was imaged using the PACT system (**a**) without and (**b**) with a piece of excised human skull in between the phantom and the light delivery path and ultrasound detection array. From the images, we conclude that the skull distorts the PA signals, thus blurring the image. However, the resultant image bears a distinct signature of the phantom being imaged. Quantitative analysis was performed by computing the contrast-to-noise ratios (CNRs) of three different regions in the reconstructed images (red box: region of interest; blue box: background region). While the image without the skull has CNRs of 392.7, 229.3 and 442.3 in the three regions, the image with the skull has CNRs of 38.9, 25.2 and 21.5 respectively. Scale bar: 5 mm.